\documentclass[10pt,twocolumn,letterpaper]{article}

\usepackage{cvpr}              % To produce the CAMERA-READY version
\usepackage{graphicx}
\usepackage{amsmath}
\usepackage{amssymb}
\usepackage{booktabs}
\usepackage{multirow}
\usepackage{ulem}
\usepackage{ulem}
\usepackage[dvipsnames]{xcolor}
\usepackage[backref]{hyperref}
\usepackage[capitalize]{cleveref}
\crefname{section}{Sec.}{Secs.}
\Crefname{section}{Section}{Sections}
\Crefname{table}{Table}{Tables}
\crefname{table}{Tab.}{Tabs.}

\def\cvprPaperID{9038} % *** Enter the CVPR Paper ID here
\def\confName{CVPR}
\def\confYear{2022}

\definecolor{citecolor}{HTML}{0071bc}
\hypersetup{
  breaklinks   = true,
  colorlinks   = true, %
  urlcolor     = RubineRed, %
  linkcolor    = RubineRed, %
}
\hypersetup{pdfnewwindow}
\newcommand{\urlNewWindow}[1]{\href[pdfnewwindow=true]{#1}{\nolinkurl{#1}}}

\begin{document}

\newcommand{\sysname}{NeuralHDHair}
\newcommand{\growname}{GrowingNet}
\newcommand{\modelname}{IRHairNet}

\definecolor{ballblue}{rgb}{0.13, 0.67, 0.8}
\definecolor{awesome}{rgb}{1.0, 0.13, 0.32}
\definecolor{asparagus}{rgb}{0.53, 0.66, 0.42}
\newcommand{\yyfs}[1]{{\color{green} \sout {#1}}}
\newcommand{\wkys}[1]{{\color{teal} \sout {#1}}}
\newcommand{\yyf}[1]{{\color{orange} #1}}
\newcommand{\yyfc}[1]{{\color{Peach} {[YF: #1]}}}

\newcommand{\wky}[1]{{\color{blue} #1}}
\newcommand{\z}[1]{{\color{red} #1}}

\newcommand{\ylc}[1]{{\color{asparagus} #1}}
\newcommand{\ylcc}[1]{{\color{awesome} #1}}
\newcommand{\ylcs}[1]{{\color{asparagus} \sout {#1}}}
\newcommand{\hb}[1]{{\color{purple}            {#1}}}
\newcommand{\hbc}[1]{{\color{cyan}            {[HB: #1]}}}

\newcommand\blfootnote[1]{%
  \begingroup
  \renewcommand\thefootnote{}\footnote{#1}%
  \addtocounter{footnote}{-1}%
  \endgroup
}

\title{\sysname: Automatic High-fidelity Hair Modeling from a Single Image Using Implicit Neural Representations}

\author{Keyu Wu$^{1*}$ 
\quad Yifan Ye$^{1*}$
\quad Lingchen Yang$^{2}$
\quad Hongbo Fu$^{3}$
\quad Kun Zhou$^{1}$
\quad Youyi Zheng$^{1\dagger}$\\[1.5mm]
$^1$ Zhejiang University \quad
$^2$ ETH Zurich \quad
$^3$ City University of Hong Kong}

\maketitle
\blfootnote{$^*$The first two authors contributed equally. The authors are affiliated with the State Key Lab of CAD\&CG. $^\dagger$Corresponding author: Youyi Zheng.}

%%%%%%%%% ABSTRACT
\begin{abstract}
   Undoubtedly, high-fidelity 3D hair plays an indispensable role in digital humans. However, existing monocular hair modeling methods are either tricky to deploy in digital systems (e.g., due to their dependence on complex user interactions or large databases) or can
   produce only a coarse geometry. In this paper, we introduce \sysname, a flexible, fully automatic system for modeling high-fidelity hair from a single image. The key enablers of our system are two carefully designed neural networks: an \modelname~(Implicit representation for hair using neural network) for inferring high-fidelity 3D hair geometric features (3D orientation field and 3D occupancy field) hierarchically and a \growname~(Growing hair strands using neural network) to efficiently generate 3D hair strands in parallel. Specifically, we perform a coarse-to-fine manner and propose a novel voxel-aligned implicit function (VIFu) to represent the global hair feature, which is further enhanced by the local details extracted from a hair luminance map. To improve the efficiency of a traditional hair growth algorithm, we adopt a local neural implicit function to grow strands based on the estimated 3D hair geometric features. Extensive experiments show that our method is capable of constructing a high-fidelity 3D hair model from a single image, both efficiently and effectively, and achieves the-state-of-the-art performance.
\end{abstract}

%%%%%%%%% BODY TEXT
\section{Introduction}

\label{sec:intro}

% introduce the problem and our work

As one of the most distinctive human characteristics, hair plays an indispensable role in digital humans \cite{cao2014face,hu2017avatar,hadap2007strands,li2015facial,luo2013structure}. Undoubtedly, a high-fidelity 3D hair model can significantly improve the realism of a virtual human. However, the existing single-view-based hair modeling methods \cite{hu2015single,chai2016autohair,zhou2018hairnet,yang2019dynamic,saito20183d} cannot sufficiently satisfy the requirements of human digitalization in terms of flexibility, simplicity, and realism. On the one hand, data-driven methods \cite{hu2015single,chai2016autohair} could achieve high-fidelity results, but are complex and not very robust, e.g., entailing a sophisticated searching and matching process based on a large hair dataset. On the other hand, deep-learning based methods \cite{zhou2018hairnet,yang2019dynamic,saito20183d} are lightweight and flexible to deploy but could only achieve coarse results.
{Thus, in this work, we consider the problem of automatic high-fidelity 3D hair modeling from a single image utilizing a learning-based method}.

\begin{figure}[t]
		\centering
		\includegraphics[width=\linewidth]{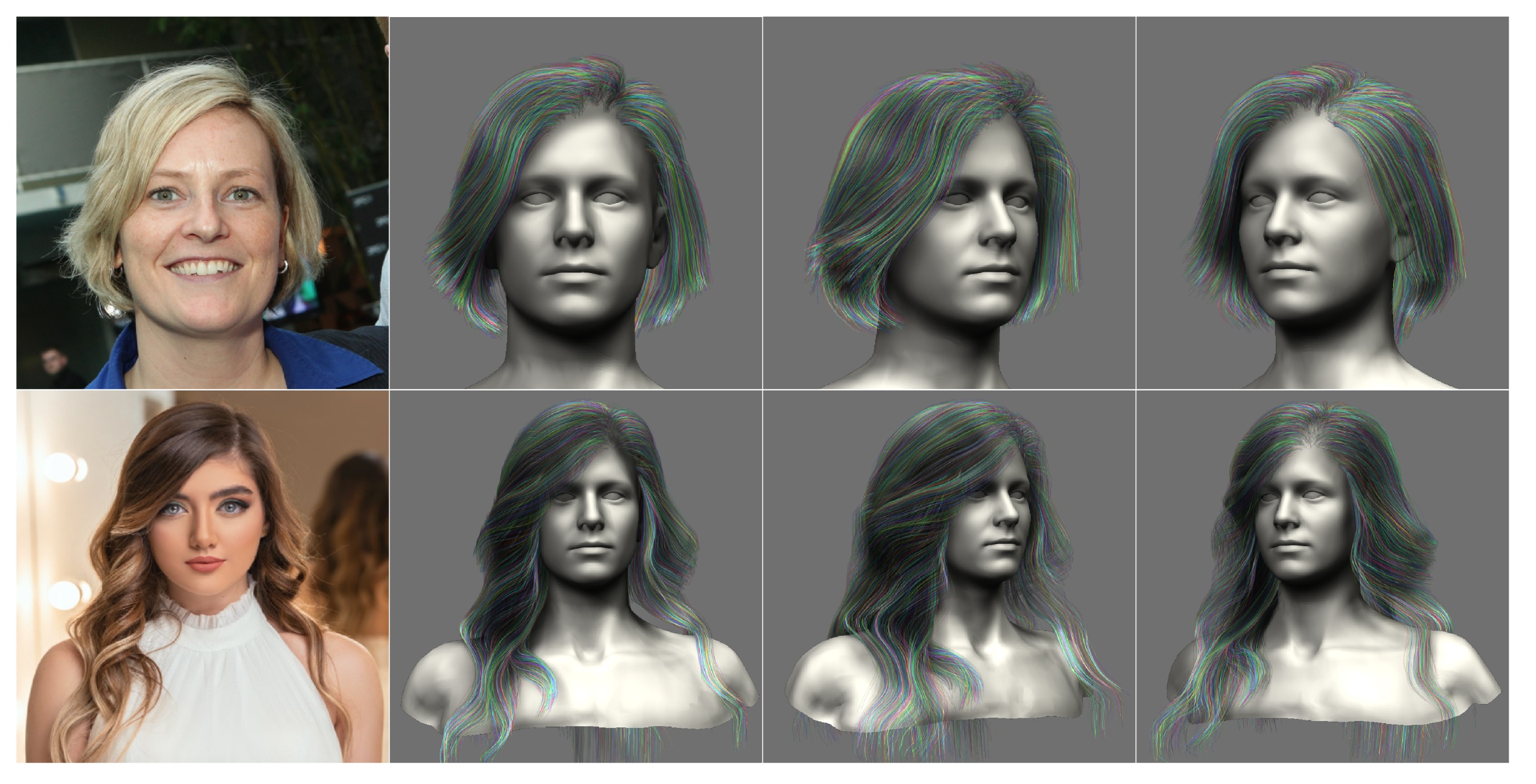}
		\caption{Given a single image, our \sysname~reconstructs a high-fidelity 3D hair model.}
		\label{fig:demo}
		\vspace{-4mm}
\end{figure}

% the challenge of the problem
%Different from 
Unlike other parts of the human body, the hair structure is more challenging to describe and extract due to remarkably intricate structures of interweaving strands, leading to extreme difficulty {in} reconstructing a high-fidelity 3D hair model only from a single view. Generally, almost all the existing methods tackle this problem in two steps: first estimating a 3D orientation field {based on} %from 
a 2D orientation map extracted from an %the 
input image and then {synthesizing} %extracting 
hair strands from the 3D orientation field. However, there are some problems with such a two-step mechanism. First, since an 2D orientation map is only a filtered version of the input hair growing information \cite{paris2004capture}, using the 2D orientation map {alone} to bridge the domain gap between real and synthetic data would unavoidably lose hair details (e.g., the relationship of occlusive strands\cite{xiao2021sketchhairsalon}). Second, the existing methods for inferring 3D orientation fields are either time-consuming
due to the use of a complex searching and matching process based on a large hair dataset \cite{chai2016autohair,hu2015single}, or liable to over-smoothness due to the use of deep networks to directly achieve image-to-voxel inference\cite{yang2019dynamic}. Third, the conventional hair growth algorithm \cite{zhang2018modeling,shen2020deepsketchhair} to extract strands from the estimated 3D orientation field are inefficient and not conducive to one-shot hair modeling. Although Zhou et al. \cite{zhou2018hairnet} have attempted to ignore the hair growth procedure by directly regressing 3D hair strands, their reconstruction results are generally unsatisfactory (see \cref{Compare}). Based on the above observations, we seek to build a fully automatic and efficient hair modeling method that can reconstruct a 3D hair model from a single image with fine-grained features (\cref{fig:demo}) while exhibiting high flexibility, e.g., reconstructing the hair model only needs one forward pass of the networks.

% Discuss possible solutions and our key idea
We found that implicit functions have excellent performance and great potential in representing \cite{park2019deepsdf} and inferring \cite{oechsle2021unisurf} 3D shapes. For example, Saito et al. \cite{saito2019pifu} introduced {PIFu} to reconstruct a whole human body from a single image, including hair. However, the quality of their hair reconstruction results is less satisfactory. We observe that unlike human body modeling, which cares only the surface geometry of a human body, 3D hair modeling needs to consider both exterior shape and interior features, which are difficult to represent by a pixel-aligned implicit function \cite{saito2020pifuhd}. To address this issue, we propose \modelname, which imposes a coarse-to-fine strategy to produce a high-fidelity %high-resolution 
3D orientation field. Specifically, we introduce a novel voxel-aligned implicit function (VIFu) to extract global information from a 2D orientation map in the coarse module. {Meanwhile, to supplement the lost local details in the 2D orientation map, we exploit a high-resolution luminance map to extract local feature and combine it with the gloabl feature in the fine module for high-fidelity hair modeling.} 

 To efficiently synthesize a hair-strand model from the 3D orientation field, we introduce \growname, a deep learning-based hair growth method by leveraging a local implicit grid representation \cite{jiang2020local}. {This is based on a key observation} that although hair geometric shapes and growing directions vary globally, they share similar features at a specific local scale. Thus, we can extract a high-level {latent code} for each local 3D orientation patch, and then train a neural implicit function ({a decoder}) to grow strands inside it based on this {latent code}. After each growing step, new local patches centered at the ends of the strands will be used to proceed with the growing. After training, it is applicable to 3D orientation fields of arbitrary resolution.

\modelname~and \growname~form the core of \sysname, a novel automatic monocular hair modeling method. We conduct extensive experiments, including comparison experiments and ablation study, and the results show that \sysname~outperforms all existing monocular hair reconstruction methods\cite{yang2019dynamic,zhou2018hairnet}. 
% our contribution

In summary, the main contributions of our work include:
	\begin{itemize}
		\item We introduce a novel fully-automatic monocular hair modeling framework, significantly outperforming the state-of-the-art methods.
		\item We introduce a coarse-to-fine hair modeling neural network (\modelname), where we use a novel voxel-aligned implicit function and a luminance map to enrich local details for high-quality hair modeling.
		\item We propose a novel hair growing network (\growname) based on a local implicit function to efficiently generate strand models with arbitrary resolution, and is an order of magnitude faster than prior methods.
	\end{itemize}

\section{Related Work}
\label{sec:Related-work}

% Hair Modeling from single image
\textbf{Hair Modeling from Single Images.} With the development of image sensing and computer graphics, 3D hair modeling \cite{paris2004capture,chai2012single,luo2013wide,hu2015single,zhang2018modeling,chai2016autohair,luo2012multi,saito20183d} has been extensively explored. Compared with multi-view-based techniques \cite{zhang2018modeling,luo2012multi}, which are typically limited to carefully regulated environments and complex hardware setups, single-view hair modeling methods show their significant strength in feasibility, generality, and efficiency. The pioneering single-view-based methods rely on different kinds of priors such as layer boundary \cite{chai2012single,chai2013dynamic} or shading cues \cite{chai2015high} to reconstruct a hair model from a single image. They often require  additional user interaction and cannot reasonably generate invisible regions. Subsequently, \cite{hu2015single,chai2016autohair} build a synthetic 3D hair database and produce impressive results from a single image based on data-driven methods. However, their reconstruction results depend highly on the quality and diversity of the database, and a large database will cause deployment difficulties. To this end, \cite{yang2019dynamic,zhou2018hairnet,saito20183d} introduce a lightweight technique for monocular hair modeling without the requirement of a pre-built database or complex hardware setups and a controlled environment. However, their methods focus on producing globally plausible models but ignore local details. In contrast, our end-to-end approach can fully automatically generate high-fidelity hair models with fine-grained features.

% implicit function
\textbf{Implicit Neural Representations.} 
Most recently, extensive studies \cite{atzmon2019controlling,park2019deepsdf,michalkiewicz2019implicit,peng2020convolutional} have been conducted for representing 3D geometry in an implicit manner due to its simplicity and effectiveness. For example, Park et al. \cite{park2019deepsdf} represent %the 
3D shapes by mapping the 3D coordinates to signed distance functions with MLPs. Mescheder et al. \cite{mescheder2019occupancy} {introduce} a new network to implicitly represent 
continuous surfaces as 3D occupancy fields for generating high-quality results at infinite resolution. However, these methods simply employ a global latent code to represent the total 3D shape, thus restricting to simple geometry and sacrificing robustness. To address {these issues, several research studies} \cite{saito2019pifu,peng2021neural,murez2020atlas,denninger20203d} {focus} %focusing 
on combining local features to represent the corresponding 3D geometry instead of directly encoding {an entire shape} into a global latent code. {PIFu proposed by Saito et al.} \cite{saito2019pifu} learns a per-pixel implicit representation from %aligning 
the pixel-aligned feature with the global context, and could preserve local details {while producing} high-fidelity reconstruction. Similarly, Jiang et al. \cite{jiang2020local} {introduce a} local implicit grid representation for arbitrary objects or scenes, {and this representation} greatly enhance{s their network's} generalization ability. {In this work, we introduce a more expressive VIFu to represent intricate hair geometry and formulate the hair growth problem as an implicit function to improve the hair growth efficiency and achieve one-shot hair modeling.}

\textbf{High-Resolution 3D Reconstruction.} 
% The urge to produce high-fidelity 3D models has led to several significant r
Several recent research focus on the reconstruction of %reconstructing 
high-quality 3D texture or geometry {based on the 2D cues}. 
For example, \cite{lazova2019360,tran2018extreme,yamaguchi2018high} estimate geometric or color details using a texture map representation. \cite{alldieck2019tex2shape,zeng2019df2net} {explore }%explored 
the unwrapped UV space by regressing displacements to improve the reconstruction quality, {and} \cite{anguelov2005scape} {deforms} % {deforms} %deformed 
the high-fidelity models of humans based on {a} %the 
data-driven method to hallucinate plausible detail{s}. 
{However, both approaches could not produce vivid details that match well with {the} truth}. 
On the other hand, \cite{saito2020pifuhd,xu2019disn} build two branches, {the coarse one and the fine one}, to {fuse} the global and local features. {Their approach} %, which
produces more detailed results while {rarely imposing extra memory burden.} %In this paper,
We adopt a {similar coarse-to-fine} strategy to reconstruct {a} high-resolution hair model.
{However, unlike their task, % with large dataset}, 
where paired training data could be easy to obtain, we lack %the
photo-realistic hair image{s paired with}} %\hbc{what kind of image? can you render the 3D model to get such an image?} 
%corresponding to the 
3D hair model{s}. 
%\hbc{\cite{saito2020pifuhd,xu2019disn} have such paired data? not clear},
{{Directly} %Thus, directly 
using {rendered images would } %the rendered image 
have a large domain gap with real image{s} (e.g.{, due to the differences in} color, {il}lumination, texture, material, etc.)}. 
We thus utilize {a} 2D orientation map as the input to our coarse branch while a luminance map {to our} %of 
fine branch. {The two maps could make up for most of the domain gap, and mutually contribute to each other.}

%%%%%%% end related work %%%%%%%
%%%%%%% end related work %%%%%%%
%%%%%%% end related work %%%%%%%
%%%%%%% end related work %%%%%%%

%%%%%% Methods start %%%%%%%
%%%%%% Methods start %%%%%%%
%%%%%% Methods start %%%%%%%
%%%%%% Methods start %%%%%%%

\begin{figure*}[t!]
		\centering
		\includegraphics[width=\textwidth]{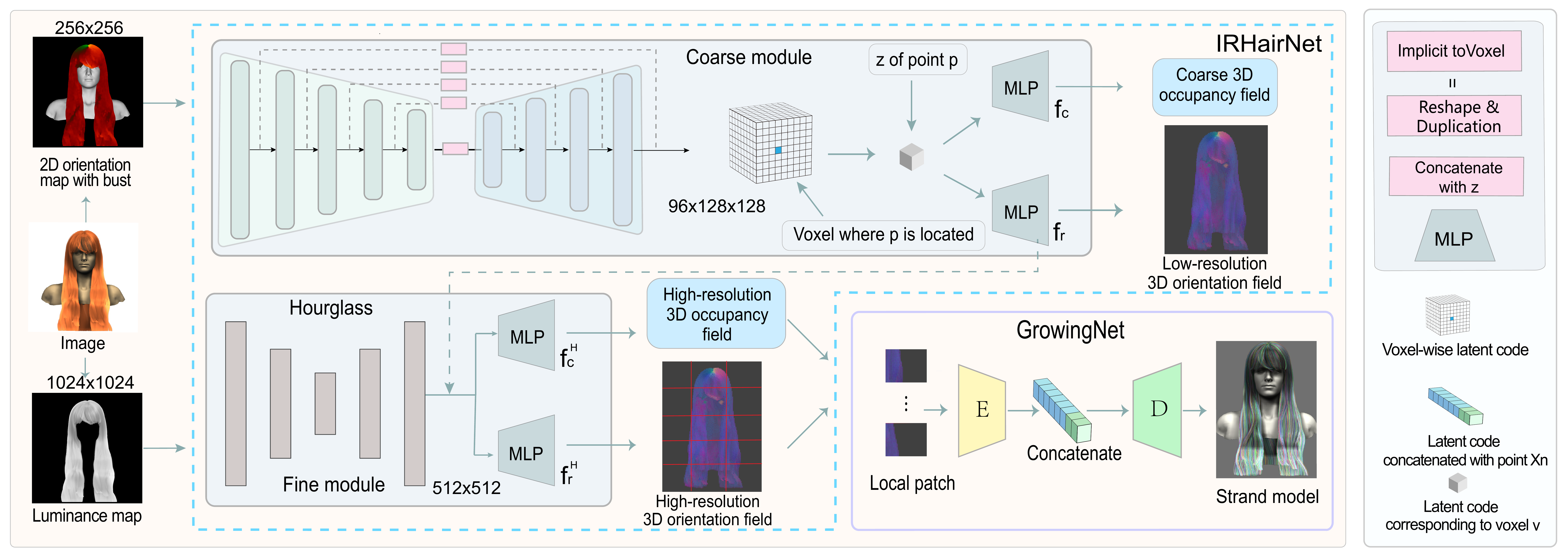}
		\caption{
% 		\hbc{Consistent capitalization needed.}
		The pipeline of \sysname. Given a single image, our \modelname~employs a voxel-aligned implicit function to predict 3D hair geometric features based on a 2D orientation map {and a high-resolution luminance map } derived from the input image. Specifically, the coarse module produces a voxel-wise latent code with global context. {The fine module extracts local details from the luminance map. We then use an MLP to decode the 3D hair geometric features from the voxel-wise {3D feature embeddings} and local details}. Afterwards, our \growname~divides the 3D orientation field into multiple patches and grow{s hair} strands in individual patches in parallel. {Finally, we remove the {out-of-boundary} hair strands using the 3D occupancy field to produce the final hair-strand model.} }
		\label{fig:pipeline}
		\vspace{-4mm}
\end{figure*}

\section{Method}
\label{sec:Method}
% Overview

\cref{fig:pipeline} shows the pipeline of \sysname. For a portrait image, we first calculate its 2D orientation map \cite{paris2004capture}, {and extract its luminance map \cite{zhang2019deep}}. Further, we automatically align {them} to the same reference bust model {to get a bust depth map \cite{yang2019dynamic}}. Then, these three maps are subsequently fed into our proposed \modelname.  {Specifically,} {it adopts a coarse-to-fine strategy and exploits a voxel-aligned implicit function} {(VIFu)} to predict the initial 3D hair geometric features (3D orientation and 3D occupancy), {which will be further enriched by the local features extracted from the hair luminance map (\cref{O2VNet}).} 
Then, the enriched 3D orientation field is divided into many patches and fed into our \growname~(\cref{Grow}) to efficiently grow {a} complete {hair-strand} model in parallel.  

\subsection{\modelname} \label{O2VNet}

%%%%  new version of the method %%%%
%%%%  new version of the method %%%%
%%%%  new version of the method %%%%

Our \modelname~is designed to generate high-resolution 3D hair geometric features from a single image. {The input to this network is comprised of a 2D orientation map, a luminance map and a fitted bust depth map, which are derived from the input portrait image. The output is a 3D orientation field, where each voxel inside contains a local growing direction, and a 3D occupancy field, where each voxel indicates whether there is any hair strand %hair 
passing through (1) or not (0). } 
% They \hbc{The 3D orientation and occupancy fields?} \wkys{respectively} \wky{both} represent hair internal features and external shape.}
% Specifically, we first obtain the 2 from the input image. Then these three maps are fed into our \modelname~, \hb{which outputs} %and then output is 
% \hb{a} %the 
% corresponding 3D occupancy field and \hb{a} 3D orientation field. \ylc{Note that, the 2D orientation map contains local 2D hair growing directions, extracted through garbor filters while the luminance map could}

\textbf{Voxel-aligned Implicit Function.} {Since the implicit function has shown impressive results in 3D shape representation,} we consider defining an implicit function to represent %the 
3D hair {geometry.} %features. % since the implicit function has shown impressive results in 3D shape representation. 
% \hbc{some of the following discussions has been repeated in the introduction. Should avoid repeated discussions.}
% \wkys{For example, Saito et al. propose\hb{d} PIFu \cite{saito2019pifu} to align local features at the pixel level to the global context and reconstruct human bodies by fitting a continuous surface. % which has achieved impressive results. However, 
% Different from human body {reconstruction focusing} on the inference of %\yyfs{exterior}
% surface features (texture and occupancy), both exterior shape and interior features are crucial for 3D hair modeling %since hair is \yyfs{not }\yyf{unlike} a flake and exhibits 
% {due to the affluent hierarchical structure of the hair}. 
% }
{As discussed in \cref{sec:intro}}, a pixel-wise implicit representation is not sufficient for representing complex 3D hair {internal} geometry and tends to
produce over-smoothed results, directly ignoring the hair's local details and spatial hierarchy (see \cref{exp}). To this end, we propose an expressive voxel-aligned implicit function (VIFu) that targets each voxel instead of each pixel to represent the intricate 3D geometry of hair. {Specifically}, given a spatial point $p$, we can obtain the local orientation value $O^{ori}_p$ and the occupancy value $O^{occ}_p$ with two implicit functions $f_{c}$ and $f_{r}$ as follows:
$f_{r}(F(v_p),p)= O^{ori} \in \mathcal{R}^3$, and $f_{c}(F(v_p),p)=O^{occ} \in \mathcal{R}^1$ , where $F(v_p)$ outputs the latent code of the voxel $v$ that $p$ locates. Since $F(v_p)$ can vary within the entire volume, it is more reasonable to represent the interior features of complex hair than PIFu \cite{saito2019pifu}. 
Furthermore, during testing, we can obtain the orientation and occupancy fields of higher resolution than during training by sampling more points inside each voxel when evaluating $f$.
% We set $O^{occ}>0.5$ as our occupancy field.
{Importantly, $F(\cdot)$ itself can be an implicit function, which maps the coordinate of the voxel $v$ to the corresponding latent code.}

% as a solution set of an implicit function $f$, e.g. $f(x)=1$, namely 3D occupancy field, similarly, we also infer the growth direction of each point as the 3D orientation field.

%reconstruct human bodies by fitting a continuous hollow surface, which has achieved impressive results. However, the 3D occupancy of the hair is generally a closed solid volume. Certainly, the 3D features inside the surface are also important for hair modeling. Therefore, their method is ill-posed for 3D hair modeling and prone to produce a smooth and continuous surface, which ignores the hair's local details and spatial hierarchy. (see \autoref{sec:exp}). We introduce a Voxel-Aligned implicit function to represent a solid volume. We define the solid volume as a solution set of an implicit function $f$, e.g. $f(x)=1$, namely 3D occupancy field. 

% Saito et al. proposed PIFU\cite{saito2019pifu} reconstruct human bodies by fitting a continuous hollow surface, which has achieved impressive results. However, the 3D occupancy of the hair is generally a closed solid surface (volume?). Certainly, the 3D features inside the surface are also important for hair modeling. Therefore, we introduce a Voxel-Aligned implicit function to represent the a solid volume. We define the solid volume as a solution set of an implicit function $f$, e.g. $f(x)=1$, namely 3D occupancy field.

% \textbf{3D Hair \hb{Feature} %\yyf{Features }%\hb{Feature} %Features 
% Representation with VIFu.} 
{\textbf{Implicit toVoxel Module with VIFu.}}
% {\hb{Now} we introduce how to obtain the feature vector $F(v)$ for each voxel.}
{Inspired by \cite{yang2019dynamic}, we use a variant U-Net architecture to achieve image-to-voxel task. The key idea is to build the skip connections between 2D and 3D features with a bridge module. Instead of simply increasing the number of 2D feature channels as in \cite{yang2019dynamic}, where the depth information is not explicitly modeled, we propose an {implicit toVoxel} module to achieve this task, which employs the idea of our VIFu.}
Specifically, this module can be formulated as:
\begin{equation}
    F(v)=\theta(I(x),Z(v)), 
\end{equation}
where for each voxel $v$, $x = \pi(v)$ is its 2D projection {coordinate}, $I(x)$ is the 2D {image feature} locating at $x$, and {$Z(v)$ is the normalized depth value}. 
% \hb{and} $I(x)$ and $\theta$ %are represent the image feature at $x$ and the MLP, respectively. 
This means each voxel's latent code $F(v)$ can be refined using its corresponding {2D feature} $I(x)$ and the normalized depth $Z(v)$ through the MLP $\theta$. 
In practice, we first duplicate the 2D features along the depth dimension of the 3D features, then apply the MLP $\theta$ based on $Z(v)$ and $I(x)$, as illustrated in \cref{fig:pipeline}. It has the following two advantages.
First, it can refine {a} pixel-wise {latent code} to {a} voxel-wise {latent code} {taking the depth into consideration, which is very important for inferring hair geometry. Second, this module can be inserted into every skip connection, thus suitable for progressive learning induced by the U-Net, e.g., the resolution of the {3D features} will gradually increase (from 6$\times$8$\times$8 to 96$\times$128$\times$128) {in} the decoding process, which helps fuse multi-level features to learn the overall shape as well as the local details.} 

{After the U-Net}, we can obtain the latent code $F(v_{p})$ at any point $p$ in the volume. Hence, the 3D hair geometric feature $O_p$ can be decoded from $F(v_p)$ as:
\begin{equation}
    O_p=f(F(v_{p}),Z(p)), 
\end{equation}
where $O_p$ represents {an} orientation or occupancy value at point $p$ and $f$ is an MLP representing $f_c$ or $f_r$ for simplicity. 
% \hbc{Is it necessary to introduce $f_c$ and $f_r$? When they will be used?}

\textbf{Coarse-to-fine Hair Modeling.} 
 Although the proposed U-Net architecture combined with VIFu {is able to achieve} %has achieved 
 decent performance (see \cref{Evaluation}), plenty of hair details present in the image still cannot be captured. %Although the proposed VIFu has achieved decent performance, plenty of hair details present in the image still cannot be captured. This is limited by the orientation map only retaining local growing directions of the hair, which lead to most local details lost.Inspired by PIFuHD\cite{saito2020pifuhd}, increasing the input resolution to provide local details may be a possible solution for high-quality reconstruction. Thus, as shown in the \autoref{fig:pipeline}, we also perform a coarse-to-fine strategy combined with the high-resolution luminance map to achieve high-resolution modeling.
 We thus utilize a coarse-to-fine strategy for high-quality reconstruction, as discussed in \cref{sec:Related-work}. For this, we design a coarse module and a fine module.
 The coarse module has the architecture of the proposed U-Net, which takes a 2D orientation map and a bust depth map as input, and outputs initial coarse 3D hair geometric features. The fine module is a Hourglass\cite{newell2016stacked} network, extracting the local features from a high-resolution luminance map. The local features will be fused into the initial hair features to achieve high-resolution modeling ultimately, as follows: 
%  \hb{As suggested by} PIFuHD \cite{saito2020pifuhd}, increasing the input resolution to accommodate {more} local details may be a possible solution for high-quality reconstruction. Thus, as shown in the Figure \ref{fig:pipeline}, we %also perform a coarse-to-fine strategy \hb{and use the orientation map together with a} % combined with the high-resolution luminance map ($1024\times1024$) to achieve high-resolution modeling. It is worth noting that the input to \hb{the} fine module is {only} a high-solution luminance map.
\begin{equation}
    O^H_p=f^H(\Omega(p),I^H(\pi(p)),Z(p)), 
\end{equation}
where $\Omega(p)$ comes from the above coarse module (the second layer output of$f$), $I^H(\pi(p))$ is the 2D fine-grained local feature extracted from the luminance map, and {$f^H$} is an MLP. 
% {Importantly, the global feature is concatenated with the local feature as well as the corresponding z-coordinates before fed into the decoder {$f^H$} to obtain the high-resolution 3D hair features.} 

Note that we convert {the input image from the RGB space} to the LAB space for obtaining a luminance map (L channel). The luminance map, highly disentangled with chrominance in the LAB space, makes it possible to describe the sophisticated hair structure. In addition, compared with the orientation map, which
only contains the growth direction {of the hair}, the luminance map can capture more local details, such as illumination and depth information, making significant improvement in realism and high-fidelity.
% Then we employ the Hourglass\cite{newell2016stacked} to capture the local features from the luminance map
% , which 
%\hbc{not clear what `which' refers to} 
% allows us to supplement %some
% {for supplementing subtle details and refining} both the final orientation and occupancy values. 
%Thus, we finally decode the high-resolution 3D features by combining the global and local features as well as the corresponding z-coordinates, utilizing an MLPs $\phi^H$. It has the following advantages: First, the luminance map is highly disentangled with chrominance and commonly used to represent the structures of the scene\cite{zhang2019deep}, which can similarly describe the hair structure. Thus, it can bridge most gaps between synthesis and real data except for materials. Second, compare with the orientation map, we can generate high-resolution luminance map to capture more local details for  high-resolution hair modeling.
 
\textbf{Loss Function.} The design of the loss function has a significant effect on the robustness of the trained model. % training a robust model.
Considering that forcing the network to fit occluded hair (especially the hair behind the head) might %may 
weaken the learning ability of the network, we assign small weights to invisible points. Similar to the previous work \cite{yang2019dynamic}, we use the Binary Cross Entropy (BCE) loss for the occupancy field and the L1 loss for the orientation field. Thus, the final loss function is given by:
\begin{equation}
    W_p=\left\{
     \begin{aligned}
    & 1  &, &Z(p) - D(p) >=\tau \\
    & 10 &, &Z(p) - D(p) <\tau \\
    \end{aligned}
    \right.
\end{equation}
\begin{equation}
\begin{split}
    \mathcal{L}_{Ori}=&\sum W_p\cdot\|O_p^{ori}-\hat{O}_p^{ori}\|^1, 
\end{split}
\end{equation}
\begin{equation}
\begin{split}
    &\mathcal{L}_{Occ}=\sum W_p\cdot(\lambda\cdot \hat{O}_p^{occ}\log (O_p^{occ})\\
    &+(1-\lambda)(1-\hat{O}_p^{occ})\log (1-(O_p^{occ}))), 
\end{split}
\end{equation}
where $D(p)$ and $Z(p)$ represent the corresponding depth value at the projected location $\pi(p)$ on the depth image $D$ and its actual z-coordinate, respectively. We regard that the points within $\tau$ ($\tau=5$) voxels under the surface of the hair are visible since the hair is not flaky. Therefore, our model can perform reasonable reconstruction for invisible hair.

\textbf{Data.} Similar to the previous research\cite{chai2016autohair,yang2019dynamic,shen2020deepsketchhair}, we collected 653 3D hair-strand models and aligned them to the same bust model within a boundary box. In addition, we also augment the data by horizontally flipping, scaling, and rotating. Then we voxelize the boundary box to prepare the training data, including 3D strand points, 3D orientation maps, and 2D orientation maps together with the luminance map. Note that we generate a luminance map from the rendered image of %image rendering by 
a 3D hair strand model at the training phase but produce it directly from an input portrait image at the testing phase. To train our \modelname, we randomly sample in the boundary box and sample points farther from the strands with Gaussian-decaying probabilities\cite{park2019deepsdf}. Finally, we perform trilinear interpolation to fill the discontinuous holes in the 3D orientation map for more robust training.

\subsection{\growname} \label{Grow}
{
Our \growname~is designed to efficiently generate a complete hair-strand model from {the 3D orientation field and 3D occupancy field estimated {by our \modelname}, where the 3D occupancy field is used to limit the growth region of hair.}

\textbf{Problem Formulation}.} We %can 
regard the each hair strand as a continuous function in a high-dimensional space similar to the spatial surface. The derivative of the spatial point on the curve represent{s} the local growing direction. If we set a fixed step size $s$, we can start from an arbitrary 3D point $x_n$ and iteratively solve the next point $x_{n+1}$ by computing the {corresponding} derivative to grow a complete strand. {Specifically,} {%Thus, 
our hair growth algorithm} can be formulated as: $x_{n+1}=x_n+ s\cdot f_g(z,x_n)$, where $z=E(O^{ori})$ represents the latent code {encoded from} the whole 3D orientation field using the encoder $E$. We attempt to apply an implicit function $f_g$ to regress the derivative of arbitrary point $x$ in the space. Then, we can produce a complete hair model by iteratively solving the next point in parallel. In fact, since $s$ is a constant, we force the network to learn the above formula and directly output the coordinates of the next point $x_{n+1}=G(z,x_n)$, where $G$ is an {MLP}. Note that the tradition hair growing algorithm entails $s$ is less than the voxel's width while in our formulation, we can change $s$ to satisfy different resolution requirements by training different MLPs.% MLPs.

\textbf{Local Implicit Hair Growing.} Since learning a global implicit function is challenging, especially for complex structures, 
we consider learning a local implicit function (a decoder) to grow hair strands. Specifically, we divide the 3D orientation field into {lots of} independent patches, which share the same decoder to grow hair strands. Based on the above considerations, we first randomly sample a 3D point in the hair volume and query its corresponding patch in the world coordinate system. Then we decode the next point according to the following formula:
\begin{equation}
    x_{n+1}=G(z_i,\frac{2}{d}(x_n-x_i)), 
\end{equation}
where $z_i$ and $x_i$ are the latent code and the center of the corresponding local patch $i$ respectively, and $d$ is the patch scale. It means that we unify all the local patches by converting the world coordinates to the local coordinate systems and normalizing the range to $[-1,1]$ before decoding.

\textbf{Hair Growth with Overlapping Latent Code.} {Although our local implicit hair growing method has an excellent performance in capturing local details, {we found that} it would produce visible artifacts in the boundary of patches (see \cref{Evaluation}). To address this issue, we employ the overlapping latent patch scheme for any two adjacent patches overlapped by half {of} the patch scale. Specifically, given an arbitrary point $x_{n}$ of a hair strand, we compute the next point $x_{n+1}$ by applying trilinear interpolation to the implicit function values of all patches overlapped at the position of $x_n$: }
%We employ the overlapping latent grid scheme, for any two adjacent grid cells overlap by half the grid-scale similar to \cite{jiang2020local}. This is because it can produce smoother hair-strand at the boundary under this strategy (see Section \ref{Evaluation}). Hence, when decoding arbitrary point $x$ in the overlapped implicit grid, the coordinate of the next point can compute by trilinear interpolation $T$ of the decoded value of each overlapped cell:
\begin{equation}
 \begin{split}
 &x_{n+1}=T(z_{\mathcal{N}},G,x_n)\\
 &=\sum_{j\in \mathcal{N}} W_j\cdot G(z_j,\frac{2}{d}(x_n-x_j)),
 \end{split}
\end{equation}
where $\mathcal{N}$ is the set of all overlapping patches at point $x_n$, and $W_j$ is the trilinear interpolation weight corresponding to the local {patch} $j$. Finally, we  generate {a} %the 
complete and smooth strand model by growing hair independently in each patch in parallel (e.g., growing 10,000 strands at the same time). 
% It is worth noting that we grow the hair-strand from the local bi-directionally to avoid the loss of cross hair strands.

\textbf{Loss Function.} We train our \growname~by minimizing the following loss function:
\begin{equation}
    \begin{split}
    \mathcal{L}=&\|T(z_{\mathcal{N}},G,x_n)-\hat{x}_{n+1}\|^1 \\
    +&\|T(z_{\mathcal{N}},G_{Inv},x_n)-\hat{x}_{n-1}\|^1, 
    \end{split}
\end{equation}
{where $G_{Inv}$ represent{s} the decoder with the same architecture as $G$, but its output is the previous point $x_{n-1}$. This is because we bi-directionally grow a hair-strand from the sampled point in the local region to avoid the loss of entangled hair strands. Thus, $\hat{x}_{n-1}$ {and} %,
$\hat{x}_{n+1}$ are the actual position{s} of the previous {and next points} %point and \hb{the} next point 
of $x_n$, respectively.
}

% $\hat{x_1}$ represents the actual position of the next point x1.

\textbf{Data.} Similarly, we use the strand model as mentioned above to train our \growname. The difference is that we control the distance between two adjacent 3D points on one strand to be approximately equal (to achieve the same step size) using B-spline interpolation for more robust training.

%%%%%% Methods end %%%%%%%
%%%%%% Methods end %%%%%%%
%%%%%% Methods end %%%%%%%
%%%%%% Methods end %%%%%%%

\begin{figure}[t]
		\centering
		\includegraphics[width=\linewidth]{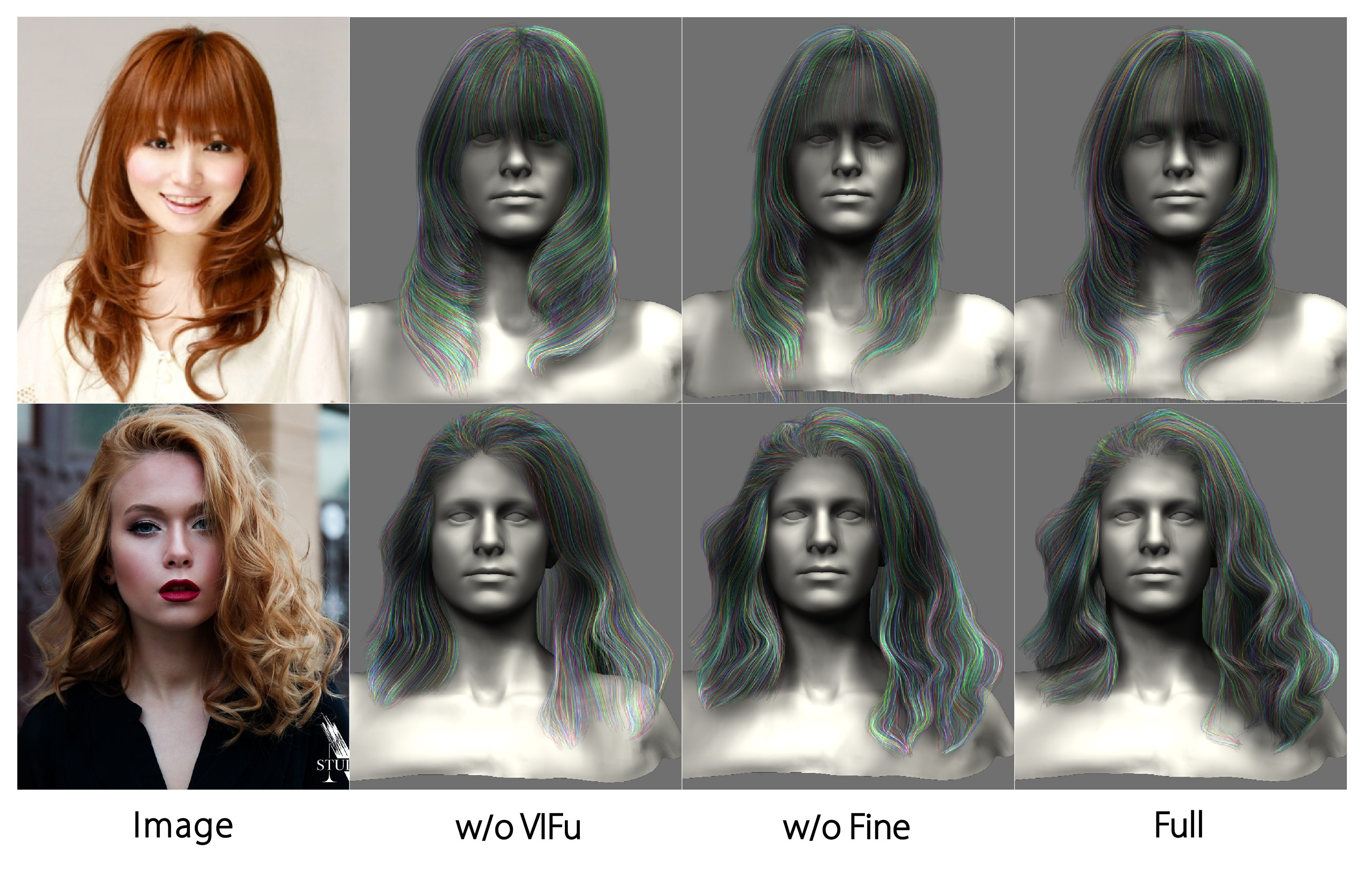}
		\caption{Qualitative evaluation of each {key} component of our \modelname. %Where 
		VIFu helps us represent complex hair structure{s}, while the fine module contributes to supplement local details.
		}
		\label{fig:ablation}
		\vspace{-4mm}
\end{figure}

\begin{figure}[ht]
		\centering
		\includegraphics[width=\linewidth]{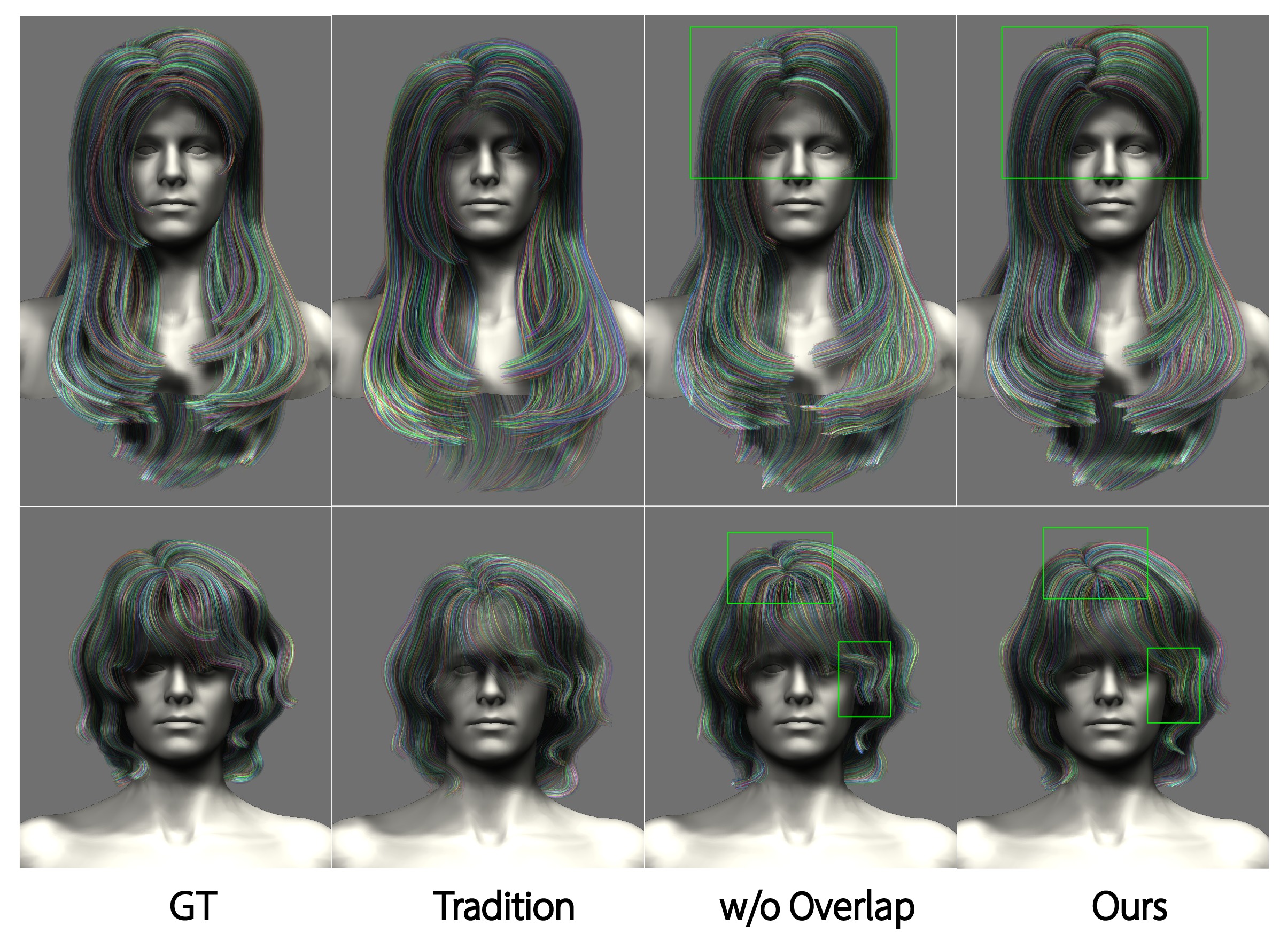}
		\caption{Qualitative evaluation of \growname. Our \growname~can efficiently achieve visually similar results to traditional methods. {The employment of overlapping latent patch scheme} helps produce smoother results across patches.
		}
		\label{fig:comparison_growing}
		\vspace{-4mm}
\end{figure}

\section{Experiments} \label{exp}
% In this section, we first introduce the implementation details of \sysname~(\cref{details}), followed by evaluating the effectiveness and necessity of each algorithmic component through {ablation studies}
% % visual and comparative experiments 
% (\cref{Evaluation}). Finally, we compare our approach with state-of-the-art methods\cite{yang2019dynamic,saito2020pifuhd,zhou2018hairnet} (\cref{Compare}).
In this section, we evaluate the effectiveness and necessity of each algorithmic component through {ablation studies} (\cref{Evaluation}). followed by comparing our approach with state-of-the-art methods\cite{yang2019dynamic,saito2020pifuhd,zhou2018hairnet,saito20183d,chai2016autohair} (\cref{Compare}). Besides, implementation details and more experimental results can be found in our supplementary materials.

\subsection{Ablation Studies} \label{Evaluation}
\textbf{Evaluation of \modelname.} To evaluate the effectiveness and necessity of each key component of our \modelname~in terms of geometric fidelity and realism, we compare our full method with two simplified settings. One simplification removes the fine module to evaluate the effect of refinement (w/o Fine) while the other removes the {implicit toVoxel} module to evaluate the significance of VIFu (w/o VIFu).

The visual comparison results are shown in \cref{fig:ablation}. When removing the overall {implicit toVoxel} module, %modules, 
our system severely suffers from the problem of over-smoothing: %, where 
the results without VIFu only contain either a plausible shape with approximate growth direction or even an incorrect structure (second row in \cref{fig:ablation}). This is because the representation ability of the network without VIFu would significantly deteriorate, making it difficult to describe intricate hair geometry. On the other hand, without the fine module, some local details are unavoidably lost, though the generated results still maintain most of the correct growth directions, especially for complex hairstyles. Employing the fine module can dramatically enhance the local details (e.g., strands' hierarchical structure) by extracting fine-grained features from the high-resolution hair luminance map while increasing the resolution of 3D hair geometric features also improve the quality of the 3D hair model (3-$th$ and 4-$th$ columns of \cref{fig:ablation}). 
% Thus, our full model is able to produce both high-fidelity and realistic results, substantiating 
The above experiments proves that the proposed VIFu has a strong representation ability for sophisticated hair geometry and such a coarse-to-fine approach can capture more local details while increasing the 3D hair model resolution.

    \begin{table}[t]
      \begin{center}

        \begin{tabular}{c|c c c c} 
            Method & Traditional & Ours& w/o Overlap  \\
          \hline
            Time(s) & 10.57 & 1.21 & 0.25\\
          \hline
        \end{tabular}
        \caption{Our \growname~is more efficient than a traditional method. Sacrificing some precision (w/o Overlap) can save a lot of time.
      }
      \vspace{-0.7cm}
        \label{fig:time_compare}
      \end{center}
      
    \end{table}

    \begin{table}[t]
      \begin{center}

        \begin{tabular}{c|c c c c} 
            Local size & 4 & 8 & 16 &32 \\
          \hline
            Time(s) & 5.75 & 1.21 &1.13&1.04\\
          \hline
        \end{tabular}
    \caption{Comparison of time consumption under different local sizes.
        }
         \vspace{-0.8cm}
        \label{fig:time_local}
      \end{center}
      
    \end{table}

\begin{figure}[b]
		\centering
		\includegraphics[width=\linewidth]{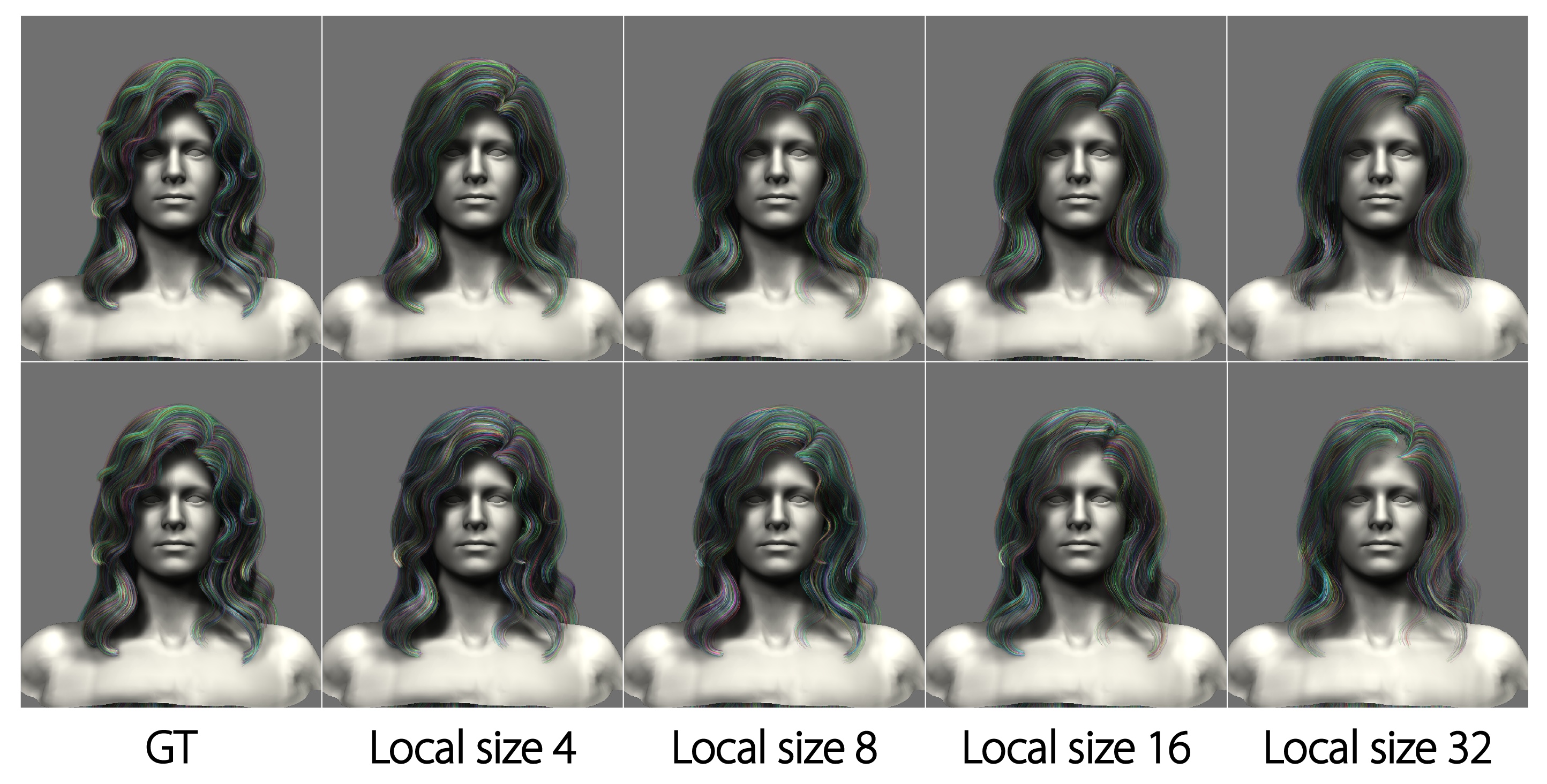}
		\caption{ Performance comparison of \growname~under different local sizes with different resolutions. The first row represents the low-resolution results ($96\times128\times128$) and the second row represents the high-resolution one ($384\times512\times512$)
		}
		\label{fig:local_size}
		\vspace{-4mm}
\end{figure}

\textbf{Evaluation of \growname.} We also qualitatively and quantitatively evaluate the effectiveness of \growname~in terms of fidelity and efficiency. We first conduct three sets of experiments on synthetic data: i) traditional hair-growth algorithm\cite{shen2020deepsketchhair}, ii) \growname~without the overlapping latent {patch} scheme, iii) our full model. \cref{fig:comparison_growing} and \cref{fig:time_compare} show our \growname~has obvious advantages over traditional hair growth algorithm in time consumption, while maintaining the same growth performance in terms of visual quality. In addition, by comparing the third and fourth columns of \cref{fig:comparison_growing}, we can see that without the overlapping latent patch scheme, the hair strands at the {patch} boundary may be discontinuous and this issue is more serious where the hair strands' growth directions change sharply. However, it is worth noting that this scheme greatly improves the efficiency at the cost of slightly reduced precision. Still, the improved efficiency is significant for the convenient and efficient application to human digitization.

% although both of them shared with the same network architecture, the lack of overlapping implicit grid strategy results in discontinuous boundary, \yyfs{On the other hand, It is worth noting that when we remove the overlapping implicit grid strategy, as we mentioned ahead, the hair strands at the grid boundary may be discontinuous, }significantly where the \yyf{hair strands }growth direction changes sharply. \yyfs{(3th and 4th column). However, its efficiency has been greatly improved, which is great significance to the convenient and efficient application to human digitization.} 

We also evaluate the reconstruction performance of \growname~on different local sizes in different resolutions ($96\times128\times128$ and $384\times512\times512$). As shown in \cref{fig:local_size}, as the input resolution increases, the quality of the result is higher. On the other hand, at the same resolution, the performance of modeling rises simultaneously with a smaller patch size, producing more accurate details, since decoding a small patch is much more simplified than a large one. It seems that we can reconstruct a high-fidelity model by simply reducing the local {patch} size. However, as shown in \cref{fig:time_local}, when the local {patch} size is reduced to less than 8, the reconstruction time increases significantly with only a tiny improvement on the modeling quality, due to its time-consuming query process similar to the traditional hair growth algorithm. 
Therefore, we choose a local {patch} size of $8\times8\times8$ to balance performance and efficiency.

\subsection{Comparisons} \label{Compare}

\begin{figure}[t]
		\centering
		\includegraphics[width=\linewidth]{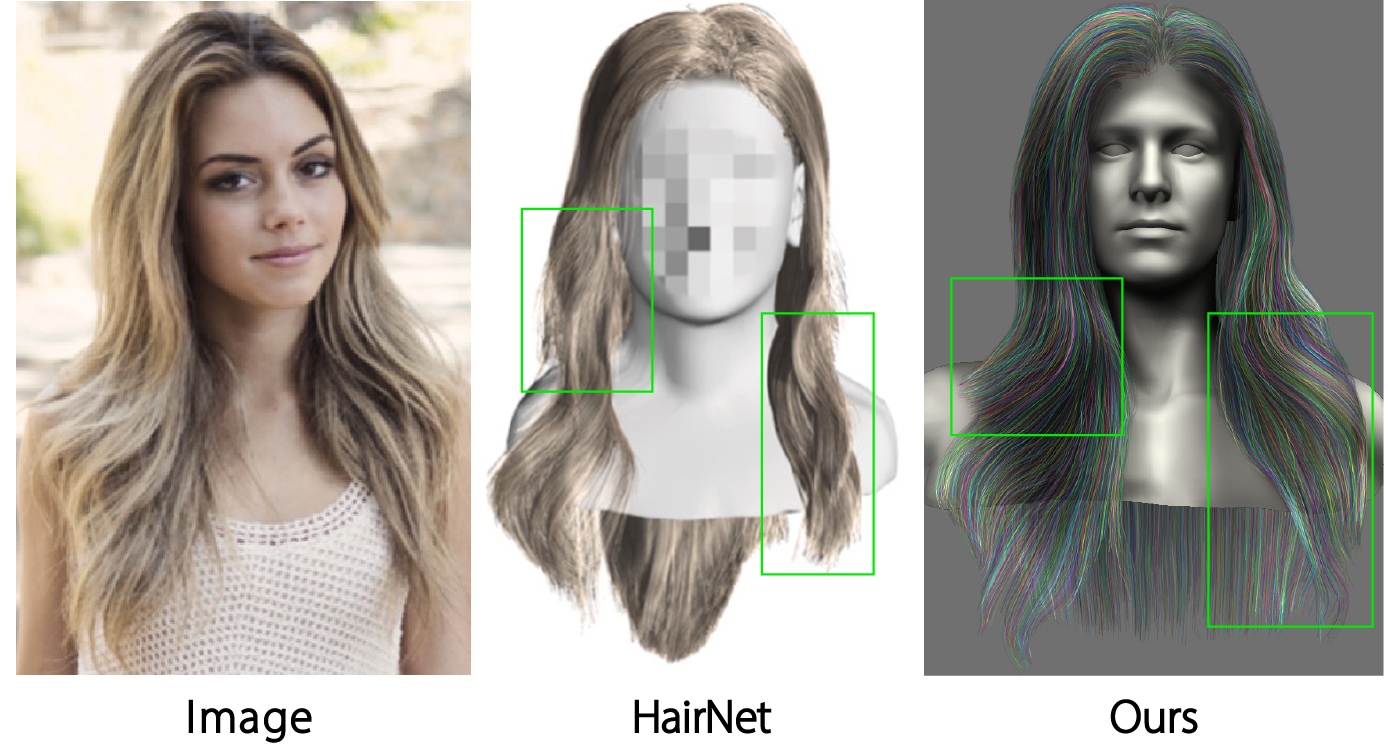}
		\caption{Compared to HairNet \cite{zhou2018hairnet}, our method leads to significantly better results in terms of shape and structure
		}
		\label{fig:compare_HairNet}
		\vspace{-4mm}
\end{figure}

\begin{figure}[t]
		\centering
		\includegraphics[width=\linewidth]{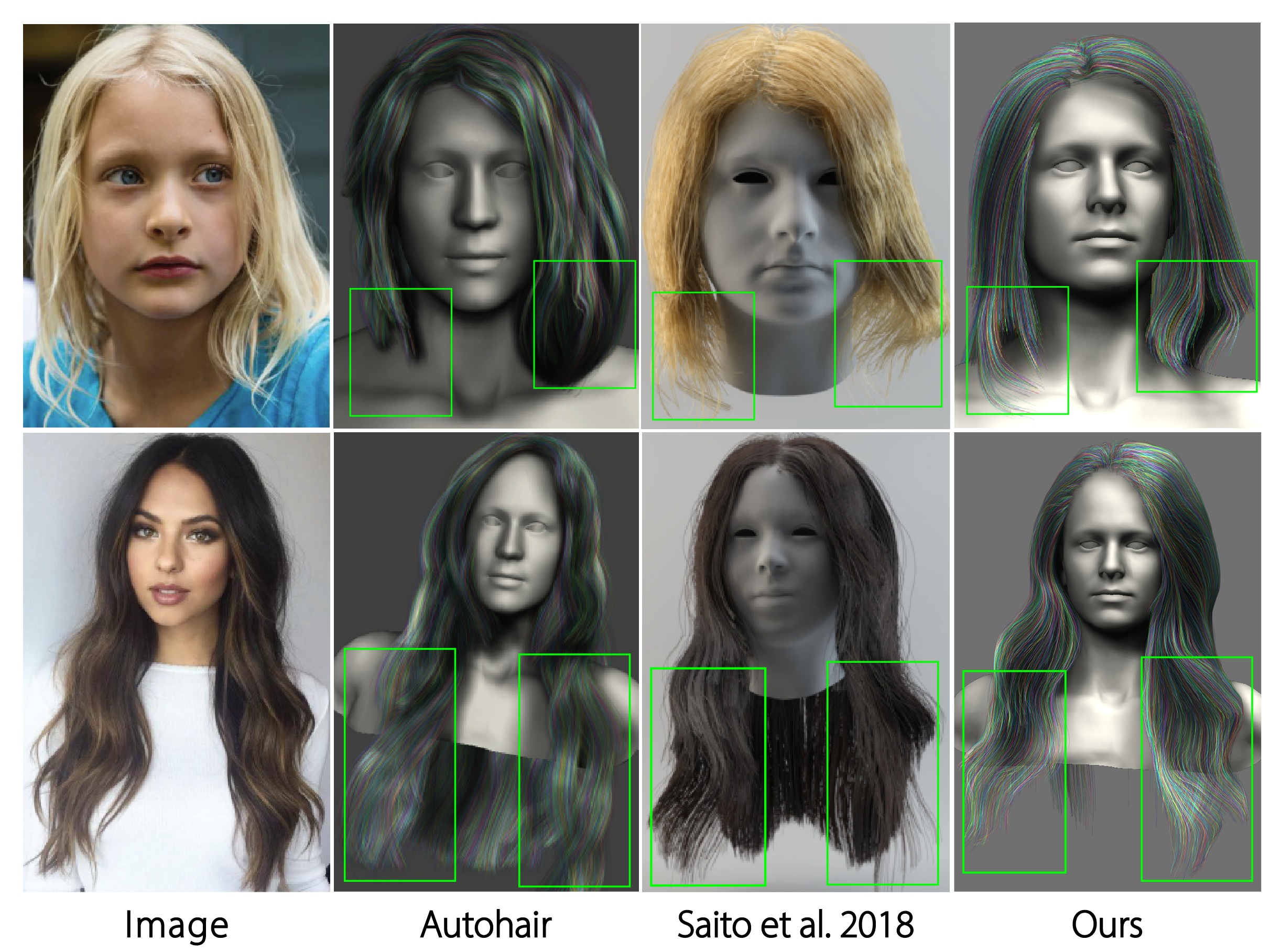}
		\caption{Comparison with Autohair\cite{chai2016autohair} and Saito et al\cite{saito20183d}. Their results are not consistent with the input image in shape or structure.
		}
		\label{fig:compare_autohair}
		\vspace{-4mm}
\end{figure}

To evaluate the effectiveness of our \sysname, we conduct several comparisons with the state-of-the-art methods \cite{zhou2018hairnet,yang2019dynamic,saito2020pifuhd,chai2016autohair,saito20183d}, where Autohair \cite{chai2016autohair} performs hair synthesis based on a data-driven approach, and HairNet \cite{zhou2018hairnet} ignores the hair growth procedure to achieve end-to-end hair modeling. In contrast, \cite{yang2019dynamic,saito20183d} perform a two-step strategy that first estimates a 3D orientation field and then synthesizes hair strands from it. PIFuHD \cite{saito2020pifuhd} is a state-of-the-art monocular high-resolution 3D modeling method based on the coarse-to-fine strategy, which can be reformulated for 3D hair modeling.

As shown in \cref{fig:compare_HairNet}, it is evident that the result by HairNet looks plausible %their result \yyf{may look plausible, }%maintain the global reasonable growth direction of hair to produce a plausible geometry, 
but the local details and even the overall shape are inconsistent with the hair in the input image. 
This is because they perform hair synthesis in a simple and crude manner by directly regressing the unordered hair strands from a single image. We also compared the reconstruction results with \cite{chai2016autohair,saito20183d}. As shown in \cref{fig:compare_autohair}, although Autohair can synthesize realistic results, its structure cannot match the input image well since the database contains limited hairstyles. On the other hand, the results of \cite{saito20183d} lack local details, and the shape cannot be consistent with the input image. In contrast, our results better maintains the global structure and local details of the hair, while ensuring the consistency of the hair shape.

\begin{figure}[t]
		\centering
		\includegraphics[width=\linewidth]{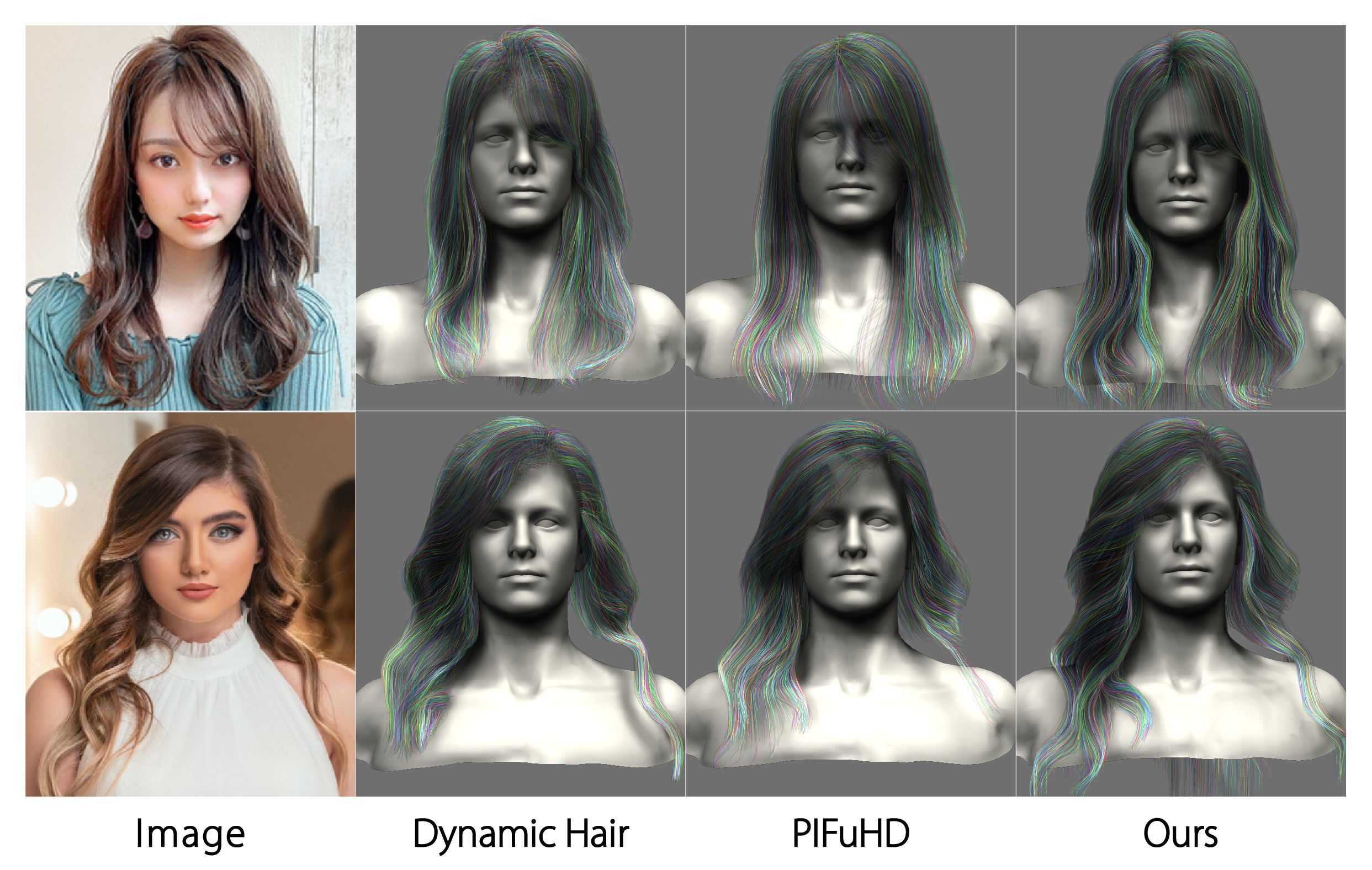}
		\caption{Comparison with PIFuHD \cite{saito2020pifuhd} and Dynamic Hair \cite{yang2019dynamic}. Compared to our method, the two existing methods cannot sufficiently describe 3D hair geometry
		and generally produce over-smooth results, especially for complicated hair structure.
		}
		\label{fig:comparison}
		\vspace{-4mm}
\end{figure}

Similar to our task, PIFuHD \cite{saito2020pifuhd} and Dynamic Hair \cite{yang2019dynamic} are committed to estimating high-fidelity 3D hair geometric features to produce realistic hair-strands models. \cref{fig:comparison} shows two representative comparison results. It can be seen that a pixel-wise implicit function adopted in PIFuHD cannot sufficiently describe the intricate hair, resulting in over-smooth results, without local details or even bad global structure. Although they attempt to supplement the local details to the coarse global features in their fine module, their generated results are still unsatisfactory due to the incorrect global structure. While \cite{yang2019dynamic} can produce more reasonable results with few details and the overall hair growth trend in their result{s} can match the input image well, many local structural details (e.g., hierarchy) cannot be captured, especially for complex hairstlyes. In contrast, with the coarse-to-fine strategy and the %\yyf{such} an
expressive implicit representation (VIFu), our method can adapt to diverse hairstyles, even extreme complex structures, and fully leverage the global features and local details to generate high-fidelity, high-resolution 3D hair models with more details.

\begin{figure}[htb]
		\centering
		\includegraphics[width=\linewidth]{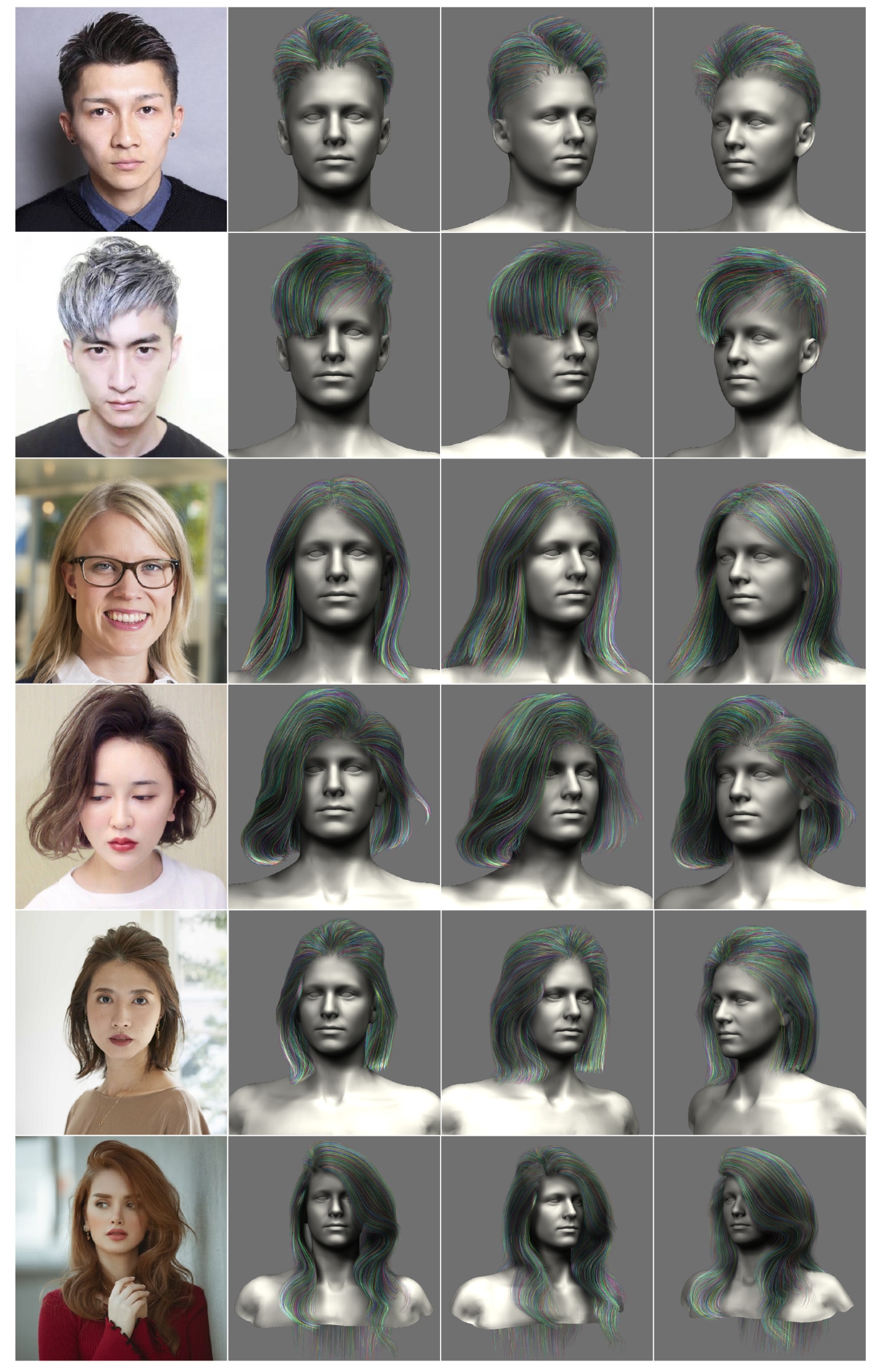}
		\caption{Fully automatic hair modeling results for various hairstyles. Our \sysname~can produce realistic, high-resolution 3D hair models with local details.}
		\label{fig:more}
		\vspace{-6mm}
\end{figure}

\section{Conclusion and Future Work}  
% We \hb{have employed} %employ 
% a coarse-to-fine strategy, and propose\hb{d} an expressive implicit representation method (VIFu) to represent global features of hair and combine its \hbc{what `its' refers to?} local details for high-resolution hair modeling. \autoref{fig:more} shows that our \sysname~can adapt to complex and diverse hairstyles and achieve high-fidelity results. Moreover, our novel hair growth algorithm can also contribute to automatic hair modeling efficiently and conveniently. Specifically, we can generate a complete hair model with only a single image as input while maintaining the state-of-the-art performance among the learning-based monocular reconstruction method\hb{s}. 
% \wky{need to rewrite!}
In this paper, we introduced a{n automatic} learning-based monocular hair modeling system \sysname~{to produce a high-fidelity hair-strand model as shown in \cref{fig:more} (Please refer to the supplemental video\footnote{\urlNewWindow{https://www.youtube.com/watch?v=g0wGRp_zKcI}} for more results).} %which can automatic produce a high-fidelity results with high-resolution and synthesize a complete hair-strand model conveniently and efficiently as shown in \autoref{fig:more}. 
{It consists of} %Our system is composed of 
two carefully designed deep leaning networks based on implicit functions: \modelname~{to provide high-resolution 3D hair geometric features. }%and \growname to , where \modelname~ provide guarantee for high-fidelity 3D hair modeling 
based on the proposed VIFu and high-resolution luminance map, and \growname~cleverly applies the local parallel growth strategy to provide an efficient hair synthesis procedure, which also supports us in producing 
a complete hair-strand model that only needs one forward pass of the networks.
We also evaluate the effectiveness and necessity of each key component of our system through ablation studies and 
demonstrate that our approach achieves state-of-the-art performance among the monocular hair modeling methods.

As far as we know, all existing hair modeling methods are limited to the quality of the 2D orientation map. Although we attempt to use the luminance map to supplement local details, we found that it must be complementary with the 2D orientation map. Utilizing only {a} luminance map as input {suffers from the problem of} overfitting due to the insufficient diversity of our 3D hair models. Thus, enriching the 3D hair dataset or adopting more realistic rendering methods is essential to reconstructing a 3D hair model directly from a single image (or a luminance map) instead of an orientation map.
% for training a robust model, which can directly reconstruct 3D hair model from a single image (or a luminance map) instead of an orientation map. 
{In addition,} for diversified portrait input, we employ landmark to automatically align it with a fixed {bust} model, which may cause the hair {not to} %cannot 
fit well with the head due to the different identities of the head. {In the future, {to significantly improve the visual quality of hair reconstruction,} we may attempt to estimate the head pose and reconstruct the corresponding 3D face while reconstructing a 3D hair model}.

\vspace{0.1cm}
{\noindent\bf Acknowledgement.}
\vspace{0.1cm} 
This work was supported in part by National Key Research \& Development Program of China (Grant No. 2018YFE0100900) and National Natural Science Foundation of China (Grant No. 62172363).

{\small
\bibliographystyle{ieee_fullname}
\normalem
\bibliography{paper}
}

\section*{Appendix}

\subsection*{A. Implementation Details}

The image encoder of the coarse module contains 5 downsampling layers and 4 upsampling layers with (32, 64, 128, 256, 256) and (256, 128, 64, 32) feature channels, respectively, where skip connections are added between them to capture more information. Note that our skip connections utilize the {implicit toVoxel} module to expand the 2D features to 3D (e.g., $8\times8$ to $6\times8\times8$, $16\times16$ to $12\times16\times16$). Finally, the size of the output voxel-wise latent code is $96\times128\times128\times64$. The MLP for the coarse module has (65, 256, 128, 64, 3) and (65, 256, 128, 64, 1) neurons for the orientation field and the occupancy field, respectively. Here the output of the second layer is concatenated with the local features as well as depth $z$ before being fed into fine module's MLP. Thus, the MLP for the fine module has (289, 512, 256, 128, 64, 3) neurons to refine the orientation field and (289, 512, 256, 128, 64, 1) neurons to refine the occupancy field. 
The coarse module is pre-trained with the 2D orientation map resized to $256\times256$ while the fine module is trained with the luminance map resized to $1024\times1024$. The \growname~is composed of an encoder and a decoder. The encoder $E$ contains several downsamplings with output channels (3, 16, 32, 64, 128) to compress the local {patch} into a latent code, and the decoder $D$ is an MLP with the number of neurons of (131, 128, 64, 32, 3). Our \modelname~and \growname~are implemented using the PyTorch framework and trained with the Adam optimizer for 2-3 days and 1 day, respectively. Our learning rate is 0.0001, and it decays every 20 epochs.

\begin{table}[b]
	\begin{center}
		\resizebox{\linewidth}{!}{
			\begin{tabular}{c|c c |c c} 
				\hline
				\multirow{2}*{Method} &\multicolumn{2}{c}{\textbf{Synthetic data}} & \multicolumn{2}{|c}{\textbf{Real data}}\\
				\cline{2-5}
				& Precision(\%) & L2 & L1 & User study(\%) \\
				\hline
				PIFuHD & 71.08 & 0.1543 & 0.2662& 14.95\\
				Dynamic Hair& 73.14& 0.1293 & 0.2091 & 19.38\\
				Ours & \textbf{76.36} & \textbf{0.1040} & \textbf{0.1458}& \textbf{65.67}\\
				
				\hline
				
			\end{tabular}
		}
		\caption{Quantitative comparison and a user study.
		}
		\label{fig:quantitative}
	\end{center}\vspace{-7mm}
\end{table}

\subsection*{B. More comparisons}
To better compare and demonstrate the effectiveness of our method, we compared with Dynamic Hair\cite{yang2019dynamic} and PIFuHD\cite{saito2020pifuhd} on large-scale test data using some quantitative metrics similar to\cite{yang2019dynamic} and conducted a user study as shown in \cref{fig:quantitative}. We use precision for occupancy field while the L2 error for orientation field on synthetic data. We calculate the L1 error between the projection of the growth direction of each point on the strand with the 2D orientation to measure the model's performance on the real data. Our user study involved 38 users and 25 test cases, and 65.67\% chose our reconstruction as the best results.

\begin{figure}[t]
	\centering
	\includegraphics[width=\linewidth]{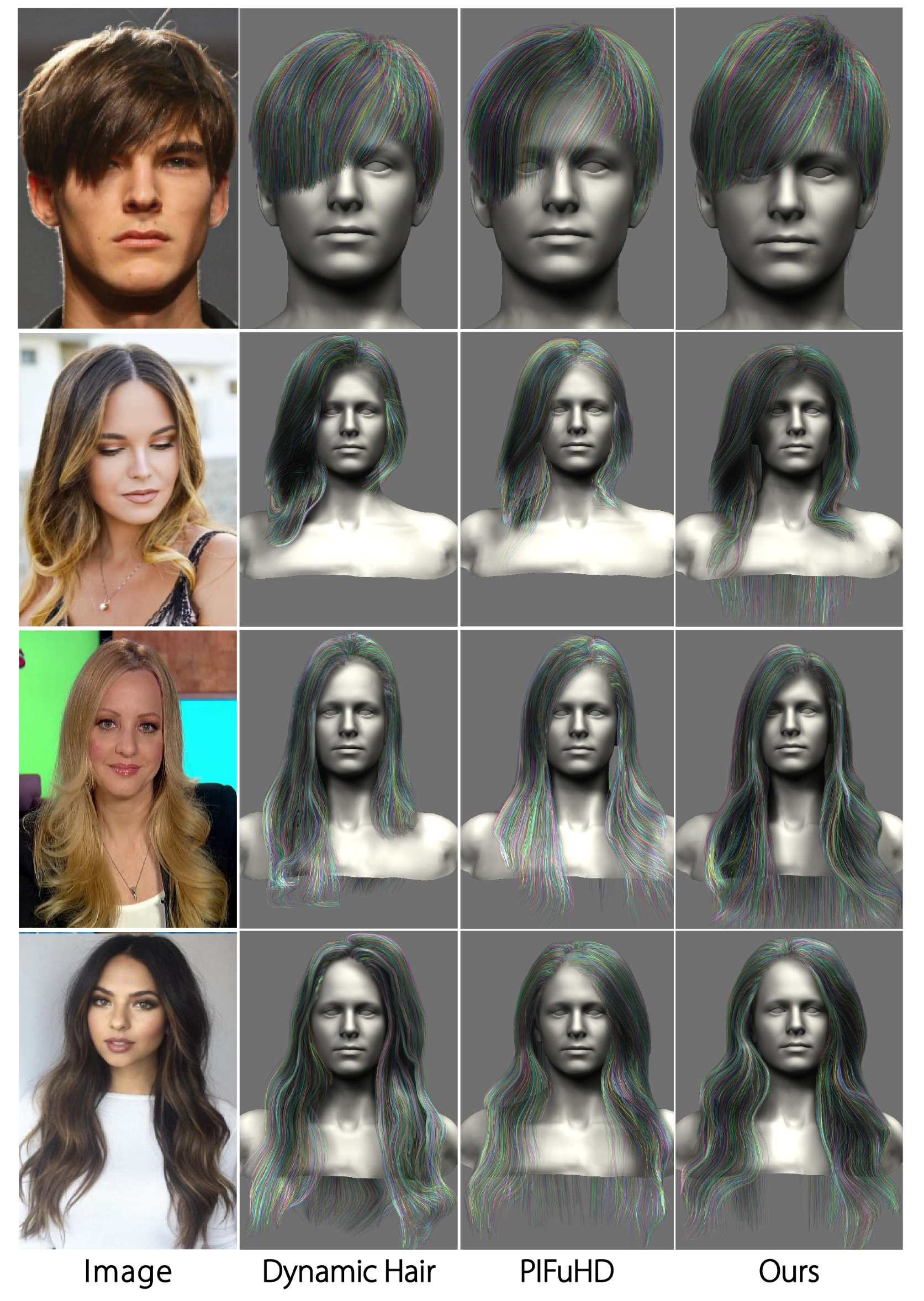}
	\caption{More qualitative comparison with Dynamic Hair\cite{yang2019dynamic} and PIFuHD\cite{saito2020pifuhd}.
	}
	\label{fig:more_compare}
	\vspace{-4mm}
\end{figure}
In addition, as shown in \cref{fig:more_compare}, we also demonstrate more qualitative comparative examples to prove that our method achieves the SOTA.

\end{document}